\documentclass{aa}
\usepackage[varg]{txfonts}
\usepackage{graphicx}
\usepackage{float}
\graphicspath{ {./plots/} }
\usepackage{natbib}
\bibpunct{(}{)}{;}{a}{}{,}
\usepackage{multicol}
\usepackage{array,multirow}
\usepackage{footnote}
\usepackage{url}
\usepackage{tabularx}
\usepackage{babel}
\usepackage{placeins}
\usepackage{hyperref}
\usepackage{cleveref}
\usepackage{orcidlink}
\begin{document}

\title{J0011+3443: a GPS compact symmetric object, gravitational lens, or dual AGN?}

\author{E.~Traianou\inst{1}\orcidlink{0000-0002-1209-6500}
 \and T.~Liu\inst{4} \orcidlink{0000-0001-5766-4287}
 \and R.~Gold\inst{1,2,3}\orcidlink{0000-0003-2492-1966}
 \and R.~Mushotzky\inst{5}\orcidlink{0000-0002-7962-5446}
}
\institute{
 Interdisziplin\"ares Zentrum f\"ur Wissenschaftliches Rechnen (IWR),
 Universit\"at Heidelberg, Im Neuenheimer Feld 205, 69120 Heidelberg, Germany\\
 \email{traianouthalia@gmail.com}
\and
Institut f\"ur Theoretische Physik, Universit\"at Heidelberg, Philosophenweg 16, 69120 Heidelberg, Germany
\and
CP3-Origins, University of Southern Denmark, Campusvej 55, 5230 Odense, Denmark
\and
 Department of Physics and Astronomy, Georgia State University,
 25 Park Place, Suite 605, Atlanta, GA 30303, USA
\and
 Department of Astronomy, University of Maryland, College Park,
 4296 Stadium Dr, College Park, MD 20742, USA
}

\date{Received [date] / Accepted [date]}

\abstract{We present new multi-frequency VLBA observations of
\object{J0011+3443} (TXS\,0008+344, $z=0.89$) at 2.3, 4.9, 8.5,
and, for the first time, 23.6\,GHz. The source consists of two compact
components A and B at a projected separation of $314\pm2$\,pc, plus
a third feature C detected at 23.6\,GHz at $0.6$\,mas from A.
Archival low-resolution radio measurements confirm an integrated
gigahertz-peaked spectrum, with an observed-frame peak frequency of
$\nu_\mathrm{peak}=0.73\pm0.08$\,GHz and a peak flux density of
$S_\mathrm{peak}=879\pm125$\,mJy. Comparison with nearly
frequency-matched low-resolution measurements shows that the VLBA
recovers $0.85\pm0.11$ of the 4.85\,GHz flux density and
$0.50\pm0.06$ of the 8.46\,GHz flux density. The lower recovered
fraction at 8.5\,GHz suggests that low-surface-brightness emission is
resolved out or falls below the VLBA surface-brightness sensitivity.
We therefore interpret the VLBA component spectra as spectra of the
compact recovered emission only. The 23.6\,GHz morphology, the
absence of a detected flat-spectrum core, the similar compact spectra
of A and B, and the steep integrated GHz spectrum favor an
interpretation of \object{J0011+3443} as a GPS-class compact symmetric
object, possibly in a short-lived or relic phase, although a dual-AGN
origin cannot be excluded without multi-epoch astrometry.}


\maketitle

\section{Introduction}
\label{sec:intro}

Gigahertz-peaked spectrum (GPS) and compact steep-spectrum (CSS) sources are radio-loud active galactic nuclei (AGN) with spectral turnovers near the GHz regime and linear sizes below $\sim$20\,kpc \citep{odea1998,odea_saikia2021}, generally interpreted as young radio galaxies still residing within the host interstellar medium \citep{begelman1996,readhead1996}, or as sources frustrated by a dense environment \citep{vanbreugel1984}. Compact symmetric objects (CSOs), with two-sided lobe or jet morphology and no extended emission on kiloparsec scales, are the parsec-to-hundred-parsec counterparts of classical doubles, with kinematic ages of $10^{3}$--$10^{4}$\,yr \citep{owsianik1998,polatidis2003}. Compact double morphologies can also arise in galaxy mergers hosting two supermassive black holes (SMBHs) \citep{begelman1980}; dual AGN at separations of tens to hundreds of parsecs are early precursors of the sub-parsec binaries targeted by pulsar timing arrays \citep[e.g.,][]{Agazie2023}. Distinguishing a dual AGN from a CSO requires very long baseline interferometry (VLBI), since both can appear as two compact
steep-spectrum components \citep[e.g.,][]{Rodriguez_2006}.

A third scenario capable of producing a compact double morphology is gravitational milli-lensing, in which a foreground mass in the range $10^6$-$10^9\,M_\odot$ splits the image of a background AGN into two components with milli-arcsecond separations \citep{wilkinson2001}. Distinguishing lensed systems from physical double sources requires multi-frequency, multi-epoch VLBI observations to test whether spectral properties and component positions are frequency-independent, as expected for a gravitational lens \citep{2025A&A...695A.169P}.
\object{J0011+3443} (TXS\,0008+344, $z=0.89$, \citealt{2010MNRAS.401.1709V}) was previously classified as a gigahertz-peaked spectrum (GPS) candidate by \citet{Panajyan1998}, based on its inverted low-frequency radio spectrum. It was later identified as a dual-AGN and milli-lens candidate in the Search for Milli-Arcsecond Lens Entities \citep[SMILE,][]{2021MNRAS.507L...6C} program, which systematically searched the Astrogeo VLBI image database \citep{petrov2021} for compact multi-component radio sources. The source consists of two compact components, hereafter A and B, separated by $\sim$39.3\,mas ($\approx314$\,pc in projection). \citet{2025A&A...695A.169P} recently presented multi-frequency EVN imaging, reporting spectral indices\footnote{We adopt the convention $S_\nu \propto \nu^{\alpha}$ throughout.} $\alpha_\mathrm{A}\approx\alpha_\mathrm{B}\approx-1.75$ between 4.3 and 7.6\,GHz and concluding that the system is more consistent with a dual-AGN or CSO nature than with gravitational lensing.

Here we present new multi-frequency VLBA observations of \object{J0011+3443} obtained in 2022, including the first 23.6\,GHz imaging, and combine them with archival low-resolution flux-density measurements to reassess the integrated radio spectrum and the compact flux recovered by VLBI. Throughout this work we adopt a flat $\Lambda$CDM cosmology with $H_0=67.4$\,km\,s$^{-1}$\,Mpc$^{-1}$ and $\Omega_m=0.315$ \citep{planck2018}.

\section{Observations and data reduction}
\label{sec:obs}

\begin{figure}
\centering
\includegraphics[width=\linewidth]{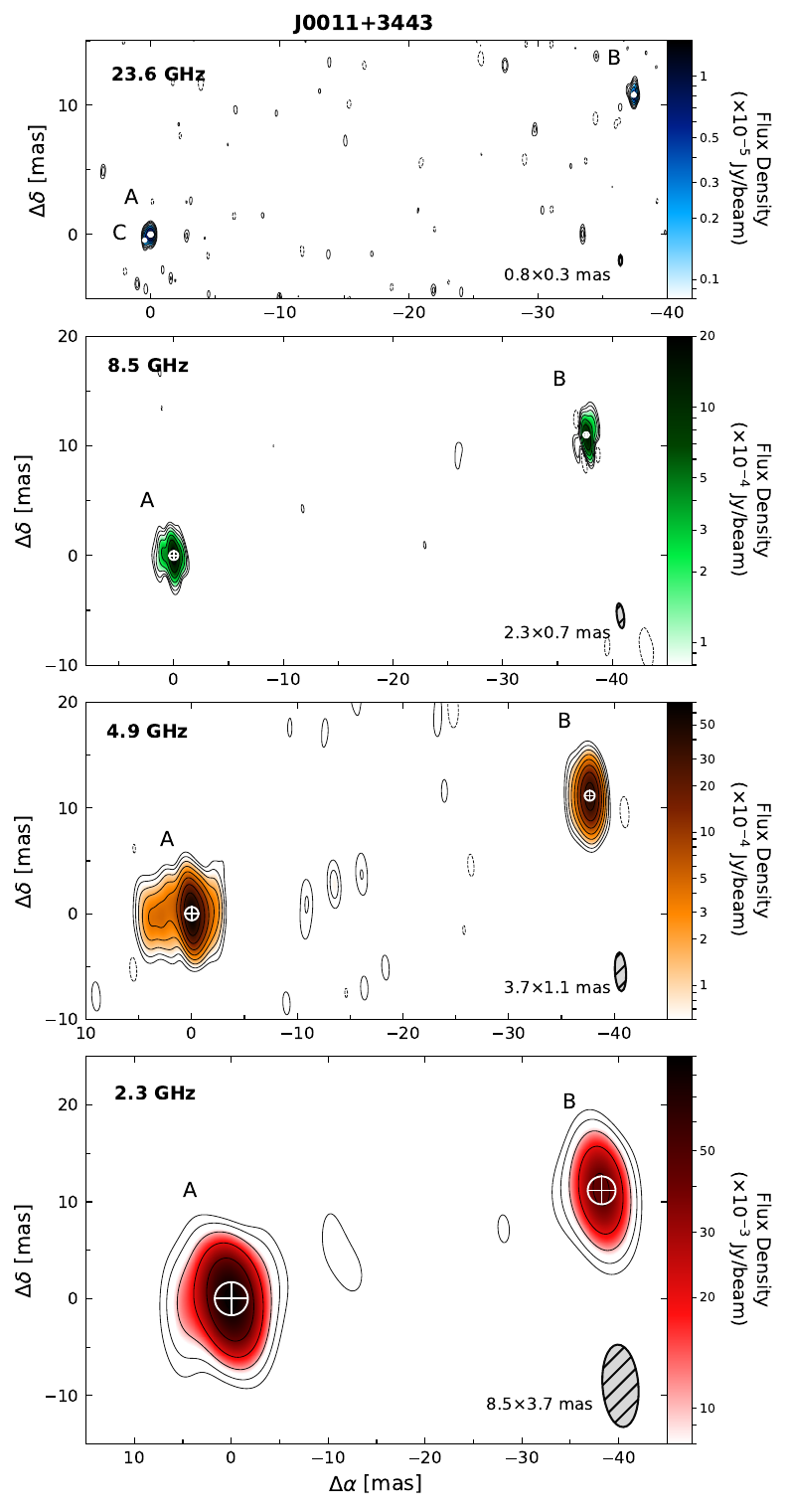}
\caption{Multi-frequency VLBA total intensity images of \object{J0011+3443} (TXS\,0008$+$344) at 23.6, 8.5, 4.9, and 2.3\,GHz (top to bottom). Contours start at ${\sim}3\sigma_\mathrm{rms}$ and increase by factors of two up to 90\% of the peak; the first negative contour is shown at $-3\sigma_\mathrm{rms}$. The off-source rms noise levels are $\sigma_\mathrm{rms}=0.05$, $0.07$, $0.04$, and $0.25$\,mJy\,beam$^{-1}$ at 23.6, 8.5, 4.9, and 2.3\,GHz, respectively. The convolving beam parameters are shown in the lower right corner of each panel. Gaussian model centroids are overlaid as white crosshairs and labeled A, B, and C.}
\label{fig:main}
\end{figure}

\object{J0011+3443} was observed with the Very Long Baseline Array (VLBA) as part of program BL292 (PI: T.~Liu), targeting lens and visible-double candidates selected by \citet{2021MNRAS.507L...6C} from the Astrogeo VLBI image database. The 2.3, 4.9, and 8.5\,GHz observations were obtained on 2022 March 8 (execution block BL292A), while the 23.6\,GHz observations were obtained on 2022 February 8 (execution block BL292G). The full sample comprises 39 sources; here we present the analysis of \object{J0011+3443}, identified as the most compelling CSO candidate and reported in advance of the full program results. The data were recorded at 4\,Gbps with the RDBE/DDC system using 2-bit sampling and dual polarization. Each intermediate-frequency (IF) channel had a bandwidth of 128\,MHz. The simultaneous 2.3/8.5\,GHz (S/X) setup comprised four IFs in total, with two IFs assigned to each band, yielding 256\,MHz of bandwidth per band. The 4.9 and 23.6\,GHz observations each used four frequency channels per polarization, yielding 512\,MHz per band. All ten VLBA stations participated. Observations were conducted in phase-referencing mode, alternating between \object{J0011+3443} and the nearby compact calibrator J0015+3216. J0237+2848 served as fringe finder.
The S/C/X and K-band execution blocks lasted approximately 1.6 and 3.2\,h, respectively, including target, phase-calibrator, and fringe-finder scans. The approximate on-source times on \object{J0011+3443} were 11.7, 11.7, 11.7, and 47.0\,min at 2.3, 4.9, 8.5, and 23.6\,GHz, respectively. The corresponding scan times on the phase calibrator J0015+3216 were 9.7, 9.8, 9.7, and 48.0\,min. The 23.6\,GHz observations used short $\sim1$\,min target-calibrator cycles to reduce atmospheric phase-transfer losses.

Data reduction followed standard VLBI procedures in \textsc{aips} \citep{greisen2003}. Ionospheric dispersive delays were corrected at 2.3, 4.9, and 8.5\,GHz using Global Ionosphere Maps in IONEX format from the International GNSS Service \citep[IGS,][]{dow2009}, retrieved from the NASA Crustal Dynamics Data Information System \citep[CDDIS,][]{noll2010,igs_ionex}. Amplitude calibration used the measured system temperatures and antenna gain curves (\textsc{aips} task \texttt{APCAL}), including an atmospheric opacity correction at all four bands, which refers the system temperatures to above-the-atmosphere values; this correction is most significant at 23.6\,GHz, where the atmospheric contribution dominates. The 23.6\,GHz flux densities are nevertheless treated conservatively: they are used to characterize the compact morphology and recovered compact emission, but not to infer a robust missing-flux fraction, because no comparable low-resolution total-flux measurement exists near this frequency. The final edited uv-coverage is shown in Appendix~\ref{app:uvcov}; the shortest projected baselines at 2.3\,GHz correspond to a largest detectable angular scale of order $\sim120$\,mas, larger than the A-B separation, although diffuse low-surface-brightness emission may still be missed because of both the surface-brightness sensitivity limits and the sparse inner uv-coverage. Channels contaminated by radio frequency interference (RFI) were flagged after exporting the IF-separated visibilities to \textsc{difmap} \citep{shepherd1994} for imaging and model fitting. \textsc{clean} deconvolution \citep{hogbom1974} followed by iterative phase and amplitude self-calibration were applied at each band. The off-source rms is consistent with the expected thermal noise at 4.9, 8.5, and 23.6\,GHz, however, at 2.3\,GHz the higher noise is dominated by residual radio-frequency interference and the limited dynamic range around the bright components rather than by the thermal limit. We then parameterized the brightness distribution by fitting circular Gaussian components to the self-calibrated data with \texttt{MODELFIT} in \textsc{Difmap} \citep{pearson1995}. Position and size uncertainties were evaluated from the local S/N \citep{1999ASPC..180..301F,2005astro.ph..3225L,2012A&A...537A..70S}; flux-density uncertainties are the quadrature sum of the local-rms statistical error and a 10\% amplitude-calibration term \citep{2009AJ....138.1874L,2024A&A...682A.154T}, which dominates for these bright components. All parameters of the fitted Gaussian components are provided in Table~\ref{tab:modelfit_txs0008} and Appendix~\ref{app:table}.

\section{Results}
\label{sec:results}

\subsection{Source morphology}
\label{sec:morph}

The VLBA images reveal a compact double morphology consistent across all four frequencies (Fig.~\ref{fig:main}). The two dominant components A and B are separated by $\sim$39.3\,mas and detected from 2.3 to 23.6\,GHz.

Component A is the brighter of the two at all frequencies. Its flux density peaks at 2.3\,GHz ($128.0\pm12.8$\,mJy) and declines steeply towards higher frequencies, reaching $1.9\pm0.2$\,mJy at 23.6\,GHz. At 23.6\,GHz a third compact feature C is detected at $0.64\pm0.06$\,mas from A at PA\,$\approx+135\degr$, with a flux density of $0.5\pm0.1$\,mJy, a $\sim$10$\sigma$ detection. Its position angle lies $\sim29\degr$ from the direction exactly opposite to B (PA\,$\approx+106\degr$), placing it on the far side of A from B and consistent with compact sub-structure associated with component A. No counterpart to C is detected near B at 23.6\,GHz. Component B follows the same spectral trend as A, declining from $67.4\pm6.7$\,mJy at 2.3\,GHz to $1.0\pm0.1$\,mJy at 23.6\,GHz, and remains unresolved or only marginally resolved at all frequencies; the full width at half maximum (FWHM) values listed for B in Table~\ref{tab:modelfit_txs0008} are upper limits set by the synthesized beam resolution.

The flux ratio $S_\mathrm{A}/S_\mathrm{B}=1.89\pm0.13$ is approximately constant across a factor of ten in frequency, indicative of strong spectral symmetry between the two components. The fitted FWHM of component A decreases monotonically with frequency, from $3.42\pm0.34$\,mas at 2.3\,GHz to $0.36\pm0.04$\,mas at 23.6\,GHz. Rather than opacity stratification of a blazar-like core, this behavior likely reflects extended low-surface-brightness hotspot or lobe emission that is progressively resolved out, or falls below the surface-brightness sensitivity, at higher frequency and angular resolution. Similar extended hotspot and lobe sub-structures are commonly observed in GPS/CSS sources \citep[e.g.,][]{AnEtal2012,Stanghellini2025}.

\subsection{Integrated radio spectrum and recovered VLBA flux density}
\label{sec:totalspec}

\begin{figure}
\centering
\includegraphics[width=\linewidth]{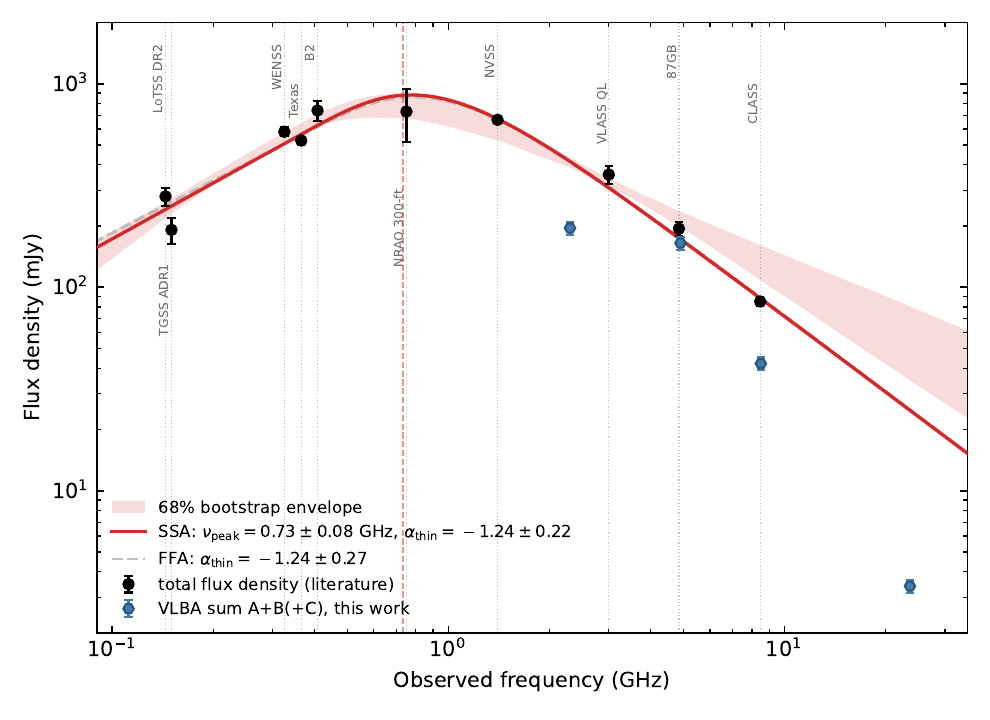}
\caption{Integrated radio spectrum of \object{J0011+3443} from
low-resolution literature measurements (Table~\ref{tab:totalspec}). The red curve shows an SSA-like absorbed-spectrum parameterization with free optically thick slope, and the gray dashed curve shows an internal FFA parameterization. The shaded region marks the 68\% bootstrap envelope of the SSA-like fit. Blue hexagons show the summed VLBA flux densities A+B, and A+B+C at 23.6\,GHz, from this work. The VLBA points are not included in the fits and are shown for comparison with the integrated low-resolution spectrum.}
\label{fig:total_spectrum}
\end{figure}

We characterize the radio spectrum of \object{J0011+3443} using integrated low-resolution flux densities from the literature, spanning 0.144-8.46\,GHz (Table~\ref{tab:totalspec}). They show a convex, GPS-like spectrum, rising from the MHz regime to a broad maximum below $\sim$1\,GHz and steepening strongly at higher frequencies. Because the measurements are non-simultaneous and heterogeneous in resolution and flux scale, we fitted two absorbed-spectrum parameterizations as phenomenological descriptions (Fig.~\ref{fig:total_spectrum}). The first is a synchrotron self-absorption (SSA) model with a free optically thick index \citep[following][]{pacholczyk1970,turler1999},

\begin{equation}
S_\nu = S_\mathrm{p}\left(\frac{\nu}{\nu_\mathrm{p}}\right)^{\alpha_\mathrm{thick}}
\frac{1-\exp\!\left[-\tau_\mathrm{m}\,(\nu/\nu_\mathrm{p})^{\,\alpha_\mathrm{thin}-\alpha_\mathrm{thick}}\right]}
{1-e^{-\tau_\mathrm{m}}},
\end{equation}
\begin{equation}
\tau_\mathrm{m} = \frac{3}{2}\left(\sqrt{1-\frac{8\alpha_\mathrm{thin}}{3\alpha_\mathrm{thick}}}-1\right),
\label{eq:ssa}
\end{equation}

where $S_\mathrm{p}$ and $\nu_\mathrm{p}$ are the peak flux density and peak frequency, $\alpha_\mathrm{thin}$ and $\alpha_\mathrm{thick}$ are the optically thin and thick spectral indices, and $\tau_\mathrm{m}$ is the optical depth at the peak. The optically thick index is left free, as the measured thick slope is much flatter than the homogeneous-source value $\alpha_\mathrm{thick}=+2.5$, as expected for an inhomogeneous double. The second is an internal free-free absorption (FFA) model \citep[e.g.,][]{tingay2003},

\begin{equation}
S_\nu = S_0\,\nu^{\alpha_\mathrm{thin}}\,\frac{1-e^{-\tau_\nu}}{\tau_\nu},
\qquad
\tau_\nu = \left(\frac{\nu}{\nu_\mathrm{t}}\right)^{-2.1},
\label{eq:ffa}
\end{equation}

where $\nu_\mathrm{t}$ is the frequency at which the free-free optical depth is unity. Both describe the data comparably well, so the present data do not determine the physical absorption mechanism. The SSA-like fit gives $\nu_\mathrm{peak}=0.73\pm0.08$\,GHz ($\nu_\mathrm{peak,rest}=1.38\pm0.15$\,GHz at $z=0.89$) and $S_\mathrm{peak}=879\pm125$\,mJy; both models yield the same optically thin index, $\alpha_\mathrm{thin}=-1.24\pm0.25$.

\begin{table*}[t]
\centering
\caption{Integrated low-resolution radio flux densities of \object{J0011+3443}.}
\label{tab:totalspec}
\begin{tabular}{lccc}
\hline\hline
Survey & $\nu$ & $S_\nu$ & Reference \\
       & (GHz) & (mJy) &  \\
\hline
LoTSS DR2   & 0.144 & $279.4\pm28.8$ & \citet{Shimwell2022} \\
TGSS ADR1   & 0.150 & $191.4\pm27.7$ & \citet{Intema2017} \\
WENSS       & 0.325 & $581.0\pm29.3$ & \citet{Rengelink1997} \\
Texas       & 0.365 & $526.0\pm23.0$ & \citet{Douglas1996} \\
B2          & 0.408 & $739.0\pm80.0$ & \citet{Colla1973} \\
NRAO 300-ft & 0.750 & $730.0\pm210.0$ & \citet{PaulinyToth1966} \\
NVSS        & 1.400 & $664.8\pm19.9$ & \citet{Condon1998} \\
VLASS QL    & 3.000 & $357.9\pm35.8$ & \citet{Gordon2021} \\
87GB        & 4.850 & $194.0\pm15.0$ & \citet{Becker1991} \\
CLASS       & 8.460 & $85.1\pm4.3$ & \citet{Jackson2007} \\
\hline
\end{tabular}
\tablefoot{The uncertainties include the catalog uncertainty and, where appropriate, an additional flux-scale term added in quadrature. The measurements are non-simultaneous and were obtained with different angular resolutions; the fits in Fig.~\ref{fig:total_spectrum} are therefore used as phenomenological descriptions of the integrated spectrum.}
\end{table*}

At frequencies for which nearly frequency-matched low-resolution measurements are available, comparison with the summed VLBA flux densities (Table~\ref{tab:recovered_flux}) shows that the VLBA recovers $0.85\pm0.11$ of the 4.85\,GHz 87GB flux density and $0.50\pm0.06$ of the 8.46\,GHz CLASS flux density. Because these measurements are non-simultaneous and were obtained at different angular resolutions, the recovered fractions should be regarded as indicative. The substantially lower recovered fraction at 8.5\,GHz nevertheless suggests that low-surface-brightness emission is resolved out or falls below the VLBA surface-brightness sensitivity. The VLBA component spectra (Table~\ref{tab:modelfit_txs0008}) consequently characterize the compact recovered emission only. Missing flux, together with differences in frequency coverage and uv-sampling, may contribute to the difference between the compact VLBA spectra and the EVN spectral indices reported
by \citet{2025A&A...695A.169P}. The VLBA component spectra are used only as a relative comparison between A and B. In this restricted sense they are still informative: both components show similarly steep compact recovered spectra and an approximately constant flux-density ratio, which argues for a common physical origin of the two radio features. However, because the absolute flux densities depend on the surface-brightness sensitivity and uv-coverage at each frequency, the compact VLBA spectra should not be used to infer the source-wide turnover frequency or the physical absorption mechanism. The steep integrated GHz spectrum is consistent with an evolved electron population (Appendix~\ref{app:aging}).

\begin{table}
\centering
\caption{VLBA recovered compact flux density compared with
nearly frequency-matched low-resolution measurements.}
\label{tab:recovered_flux}
\begin{tabular}{cccc}
\hline\hline
$\nu_\mathrm{VLBA}$ & $S_\mathrm{VLBA}$ & $S_\mathrm{low-res}$ & $f_\mathrm{rec}$ \\
(GHz) & (mJy) & (mJy) & \\
\hline
4.9 & $164.8\pm16.5$ & $194.0\pm15.0$ & $0.85\pm0.11$ \\
8.5 & $42.2\pm4.2$   & $85.1\pm4.3$   & $0.50\pm0.06$ \\
\hline
\end{tabular}
\tablefoot{$S_\mathrm{VLBA}$ is the summed compact flux density of components A and B. $S_\mathrm{low-res}$ is the nearly frequency-matched low-resolution flux density: the 87GB measurement at 4.85\,GHz \citep{Becker1991} and the CLASS measurement at 8.46\,GHz \citep{Jackson2007}, respectively. The uncertainties on $f_\mathrm{rec}=S_\mathrm{VLBA}/S_\mathrm{low-res}$ were obtained by propagating the VLBA and catalog flux-density uncertainties. Because the VLBA and catalog measurements are non-simultaneous and were obtained at different angular resolutions, the recovered fractions should be regarded as indicative.}
\end{table}

\subsection{Brightness temperatures and frequency-dependent separation}
\label{sec:tb}

Source-frame brightness temperatures were computed as

\begin{equation}
T_\mathrm{b} = 1.22\times10^{12}\,(1+z)\,
\frac{S\,[\mathrm{Jy}]}{\nu^2\,[\mathrm{GHz}]\;\theta^2\,[\mathrm{mas}]}
\;\;\mathrm{K},
\label{eq:tb}
\end{equation}

\noindent where $\theta$ is the Gaussian FWHM \citep[e.g.,][]{2025A&A...700A..16T}. For unresolved modelfit components, we set $\theta_\mathrm{obs} = \theta_\mathrm{min}$ \citep{2005astro.ph..3225L} of each knot, and consider the resulting estimate of $T_\mathrm{b}$ as a lower limit. All values are listed in Table~\ref{tab:modelfit_txs0008}.

Knot A reaches a peak brightness temperature of $T_\mathrm{b,A}=6.3\times10^{9}$\,K at 4.9\,GHz. This value is roughly one order of magnitude below the equipartition brightness temperature $T_\mathrm{b,eq}\approx5\times10^{10}$\,K \citep{readhead1994}, consistent
with synchrotron emission from compact lobe or hotspot plasma rather than a Doppler-boosted core. The lower limit on $T_\mathrm{b,B}$ at 4.9\,GHz is $\gtrsim6.2\times10^{9}$\,K, approximately consistent with the value for A, although B is unresolved or only marginally resolved and this comparison is therefore indicative rather than precise. Knot C has a lower limit $T_\mathrm{b,C}\gtrsim3\times10^{7}$\,K at 23.6\,GHz.

The projected separation between A and B shows no statistically significant frequency dependence. The largest difference in the A-B separation between any pair of frequencies is $0.90\pm1.68$\,mas across 2.3-23.6\,GHz, or $0.35\pm1.66$\,mas when restricted to 4.9-23.6\,GHz, in both cases consistent with no frequency dependence. These values are consistent with the EVN separations of $39.38\pm0.19$ and $39.40\pm0.17$\,mas reported by \citet{2025A&A...695A.169P} at 4.3 and 4.9\,GHz in 2016 and 2020, demonstrating stability over six years.

\section{Discussion}
\label{sec:discussion}

Our multi-frequency VLBA data allow us to assess three competing interpretations for the compact double morphology of \object{J0011+3443}.

\subsection{Gravitational lensing}
\label{sec:lens}

The approximately constant compact flux-density ratio, together with the similar compact recovered spectra of A and B (Table~\ref{tab:modelfit_txs0008}), could at first sight be consistent with a lensed system in which two images of a single background source are observed. However, several observables disfavor this interpretation. Our 23.6\,GHz VLBA imaging resolves component A into a compact double structure while B remains unresolved or only marginally resolved. Gravitational lensing preserves surface brightness modulo magnification and parity, so a simple lens would be expected to produce broadly similar sub-structure in both images; the observed asymmetry is therefore difficult to reconcile with standard lensing expectations. This asymmetry is confirmed by archival EVN imaging \citep{2025A&A...695A.169P}, which resolves A further into three sub-components (A0, A1, A2) while B remains compact. The radio source is identified with \object{B2 0008+34} at $z=0.89$, consistent with this galaxy being the host of the active radio source. No independently identified lower-redshift luminous deflector is known along the line of sight, providing no positive evidence for a conventional luminous lens. This does not, however, exclude milli-lensing by a dark or unresolved compact foreground mass. Taken together, these arguments disfavor, but do not definitively rule out, a lensing origin.

\subsection{Compact symmetric object}
\label{sec:cso}

The integrated low-resolution spectrum peaks at $\nu_\mathrm{peak,rest}=1.38\pm0.15$\,GHz, confirming the GPS classification independently of the VLBA component spectra. The projected size of $\approx314$\,pc is consistent with the compact sizes of GPS/CSS sources \citep{odea1998,odea_saikia2021}. For its 0.31\,kpc size, the turnover-size anticorrelation of \citet{1997AJ....113..148O} predicts a rest-frame peak of $\approx$1.3\,GHz, in excellent agreement with our measurement. In this picture, A and B are the two hotspot or lobe complexes of a young jet still residing within the host galaxy, fed by a central engine located between them. The central engine is undetected at all four frequencies, including 23.6\,GHz, consistent with being faint, absorbed, or below the current VLBA sensitivity.

The compact sub-structure of A, including component C in our 23.6\,GHz image and A0, A1, and A2 in the EVN images of \citet{2025A&A...695A.169P}, is naturally associated with a jet-termination or hotspot region. The steep integrated GHz spectrum, together with the steep compact recovered spectra of A and B, is consistent with evolved lobe plasma. Under the hotspot or lobe interpretation, we estimated the equipartition magnetic field of the compact VLBA components using the 4.9-GHz flux densities and sizes, adopting a filling factor of unity, equal energy in relativistic protons and electrons, and no Doppler boosting. This gives
$B_\mathrm{eq}\approx11$-$12$\,mG (Appendix~\ref{app:Beq}), consistent with values found in compact young radio sources and in the hotspots of CSOs and high-frequency peakers (HFPs)
\citep[e.g.,][]{2008A&A...487..885O,Nyland_2026}. These estimates characterize the compact emission under the adopted assumptions; they do not provide a radiative age.

The non-detection of a central engine and the steep GHz spectrum are also consistent with a short-lived or fading CSO scenario, in which intermittent central-engine activity may strongly influence the subsequent radio morphology and evolution
\citep[e.g.,][]{Gugliucci2005,KunertBajraszewska2005,Orienti2010,An2026,Kiehlmann2024}. In such systems, jet activity may stop before the source grows into a large-scale radio galaxy, leaving compact lobes or hotspots that fade rapidly through radiative and adiabatic losses. Young radio remnants are rare, but similar interpretations have been discussed in the context of intermittent radio-source activity
\citep[e.g.,][]{KunertBajraszewska2010,Orienti2012,Morganti2017,Orienti2023}. In this picture, the central engine in \object{J0011+3443} may have produced the two compact lobe complexes before the supply of fresh particles declined. Component C could then represent compact sub-structure or recently injected plasma associated with the A-side hotspot. This scenario remains speculative, but is testable with multi-epoch 23.6\,GHz VLBA imaging: a traveling feature should show measurable proper motion, while fading lobe plasma should show a monotonic decline in flux density.

We also note that an alternative morphological interpretation cannot be entirely excluded: component A could harbor an active nucleus while component B represents a single hotspot in the extended emission. However, the similar compact spectra, approximately constant flux-density ratio, and comparable brightness temperatures of A and B argue against such a strongly asymmetric configuration, which would be expected to produce a flat or inverted spectrum in the core component with no spectral counterpart in B.

\subsection{Dual AGN}
\label{sec:dual}

The sub-structure of component A could in principle indicate an active nucleus, implying that both components host independent SMBHs. However, this scenario is less natural than the CSO interpretation and would require a coincidence of physical conditions. Both components exhibit steep compact recovered spectra and sub-equipartition brightness temperatures, which are more characteristic of lobe or hotspot plasma than of compact self-absorbed AGN cores. Their similar compact spectra and approximately constant flux-density ratio would also be difficult to explain if A and B were two unrelated active nuclei. No flat-spectrum core is detected at 23.6\,GHz in either component. Moreover, the source is unpolarized at 8.4\,GHz ($p\lesssim1$\%; \citealt{Jackson2007}), characteristic of CSO lobes embedded in dense ionized media rather than of blazar-like cores. We also note that the SMILE selection criterion \citep{2021MNRAS.507L...6C} requires flat or inverted spectra ($\alpha\gtrsim-0.5$) for dual-AGN candidates; \object{J0011+3443} therefore falls outside this criterion, reinforcing the importance of broadband spectral and morphological follow-up in future searches for sub-kiloparsec dual AGN.

\section{Conclusions}
\label{sec:conclusions}

We have presented the first four-frequency VLBA study of \object{J0011+3443} at 2.3, 4.9, 8.5, and 23.6\,GHz.

\begin{enumerate}

\item The source consists of two compact components A and B at a projected separation of $39.3\pm0.3$\,mas ($314\pm2$\,pc) at PA\,$\approx-74\degr$, stable over six years of multi-epoch data. A third compact feature C is detected at 23.6\,GHz at $0.6$\,mas from A on the side opposite to B.

\item Archival low-resolution measurements confirm an integrated GPS spectrum with $\nu_\mathrm{peak}=0.73\pm0.08$\,GHz ($1.38\pm0.15$\,GHz rest frame) and $S_\mathrm{peak}=879\pm125$\,mJy; SSA-like and FFA-like parameterizations describe it equally well. The VLBA recovers $0.85\pm0.11$ of the nearly frequency-matched 4.85\,GHz 87GB flux density and $0.50\pm0.06$ of the 8.46\,GHz CLASS flux density, suggesting significant missing low-surface-brightness emission, particularly at 8.5\,GHz. The component spectra are therefore interpreted as compact recovered emission only, and this missing flux likely contributes to the discrepancy with the EVN index of \citet{2025A&A...695A.169P}.

\item Component A is resolved and reaches a peak brightness temperature $T_\mathrm{b,A}=6.3\times10^{9}$\,K, sub-equipartition by roughly one order of magnitude and characteristic of compact lobe or hotspot plasma. Component B is unresolved, with a brightness-temperature lower limit of $T_\mathrm{b,B}\gtrsim6.2\times10^{9}$\,K.

\item The 23.6\,GHz morphology, the absence of a detected flat-spectrum core, the similar compact spectra and flux-density ratio of A and B, and the steep integrated GHz spectrum favor a GPS-class CSO interpretation, possibly in a short-lived or relic phase. A dual-AGN origin cannot be definitively excluded without multi-epoch astrometry.

\end{enumerate}

Multi-epoch 23.6\,GHz VLBA astrometry to search for proper motion of knot C, combined with deeper optical and near-infrared imaging to constrain any foreground galaxy along the line of sight, would place substantially stronger constraints on the nature of this system.

\begin{acknowledgements}
The authors thank Thomas Greulich for his invaluable support in setting up the server used for this analysis. This work has received funding from the European Union's Horizon Europe research and innovation program under grant agreement No.\,101093934 (RadioBlocks). The National Radio
Astronomy Observatory is a facility of the National Science Foundation operated under cooperative agreement by Associated Universities, Inc.
\end{acknowledgements}

\bibliographystyle{aa}
\bibliography{aanda}

@article{odea1998,
  author  = {{O'Dea}, C.~P.},
  title   = {The Compact Steep-Spectrum and Gigahertz-Peaked-Spectrum Radio Sources},
  journal = {\pasp},
  year    = {1998},
  volume  = {110},
  pages   = {493--532},
  doi     = {10.1086/316162}
}

@article{odea_saikia2021,
  author  = {{O'Dea}, C.~P. and {Saikia}, D.~J.},
  title   = {Compact steep-spectrum and peaked-spectrum radio sources},
  journal = {\aapr},
  year    = {2021},
  volume  = {29},
  pages   = {3},
  doi     = {10.1007/s00159-021-00131-w}
}

@incollection{begelman1996,
  author    = {{Begelman}, M.~C.},
  title     = {Baby Cygnus {A}'s},
  booktitle = {Cygnus A -- Study of a Radio Galaxy},
  year      = {1996},
  editor    = {{Carilli}, C.~L. and {Harris}, D.~E.},
  pages     = {209},
  publisher = {Cambridge University Press}
}

@article{readhead1996,
  author  = {{Readhead}, A.~C.~S. and {Taylor}, G.~B. and {Xu}, W. and
             {Pearson}, T.~J. and {Wilkinson}, P.~N. and {Polatidis}, A.~G.},
  title   = {The Statistics and Ages of Compact Symmetric Objects},
  journal = {\apj},
  year    = {1996},
  volume  = {460},
  pages   = {612--633},
  doi     = {10.1086/177003}
}

@article{owsianik1998,
  author  = {{Owsianik}, I. and {Conway}, J.~E.},
  title   = {First detection of apparent superluminal motion in the compact
             symmetric object {OQ}\,208},
  journal = {\aap},
  year    = {1998},
  volume  = {337},
  pages   = {69--78}
}

@article{polatidis2003,
  author  = {{Polatidis}, A.~G. and {Conway}, J.~E.},
  title   = {Proper Motions in Compact Symmetric Objects},
  journal = {\pasa},
  year    = {2003},
  volume  = {20},
  pages   = {69--74},
  doi     = {10.1071/AS02053}
}

@article{begelman1980,
  author  = {{Begelman}, M.~C. and {Blandford}, R.~D. and {Rees}, M.~J.},
  title   = {Massive black hole binaries in active galactic nuclei},
  journal = {Nature},
  year    = {1980},
  volume  = {287},
  pages   = {307--309},
  doi     = {10.1038/287307a0}
}

@article{turler1999,
  author  = {T{\"u}rler, M. and Courvoisier, T.~J.~L. and Paltani, S.},
  title   = {Modelling the submillimetre-to-radio flares of {3C~273}},
  journal = {\aap},
  year    = {1999},
  volume  = {349},
  pages   = {45--56}
}

@book{pacholczyk1970,
  author    = {Pacho{\l}czyk, A.~G.},
  title     = {Radio Astrophysics. Nonthermal Processes in Galactic
               and Extragalactic Sources},
  publisher = {W.~H. Freeman},
  address   = {San Francisco},
  year      = {1970}
}

@ARTICLE{2024A&A...682A.154T,
       author = {{Traianou}, Efthalia and {Krichbaum}, Thomas P. and {G{\'o}mez}, Jos{\'e} L. and {Lico}, Rocco and {Paraschos}, Georgios Filippos and {Cho}, Ilje and {Ros}, Eduardo and {Zhao}, Guang-Yao and {Liodakis}, Ioannis and {Dahale}, Rohan and {Toscano}, Teresa and {Fuentes}, Antonio and {Foschi}, Marianna and {Casadio}, Carolina and {MacDonald}, Nicholas and {Kim}, Jae-Young and {Hervet}, Olivier and {Jorstad}, Svetlana and {Lobanov}, Andrei P. and {Hodgson}, Jeffrey and {Myserlis}, Ioannis and {Agudo}, Ivan and {Zensus}, Anton J. and {Marscher}, Alan P.},
        title = "{Lost in the curve: Investigating the disappearing knots in blazar 3C 454.3}",
      journal = {\aap},
         year = 2024,
        month = feb,
       volume = {682},
          eid = {A154},
        pages = {A154},
          doi = {10.1051/0004-6361/202347267},
archivePrefix = {arXiv},
       eprint = {2312.15556},
 primaryClass = {astro-ph.HE}
}

@ARTICLE{tingay2003,
  author  = {{Tingay}, S.~J. and {de Kool}, M.},
  title   = {{Compact Symmetric Objects and Radio Galaxy Evolution}},
  journal = {\aj},
  year    = 2003,
  volume  = {126},
  number  = {2},
  pages   = {723},
  doi     = {10.1086/375648}
}

@article{wilkinson2001,
  author  = {Wilkinson, P.~N. and Henstock, D.~R. and Browne, I.~W.
             and Polatidis, A.~G. and Augusto, P. and Readhead,
             A.~C.~S. and Xu, W. and Taylor, G.~B. and Pearson,
             T.~J. and Vermeulen, R.~C.},
  title   = {Limits on the Cosmological Abundance of Supermassive
             Compact Objects from a Search for Multiple Imaging in
             Compact Radio Sources},
  journal = {Phys.~Rev.~Lett.},
  year    = {2001},
  volume  = {86},
  pages   = {584--587},
  doi     = {10.1103/PhysRevLett.86.584}
}

@ARTICLE{readhead1994,
       author = {{Readhead}, Anthony C.~S.},
        title = "{Equipartition Brightness Temperature and the Inverse Compton Catastrophe}",
      journal = {\apj},
         year = 1994,
        month = may,
       volume = {426},
        pages = {51},
          doi = {10.1086/174038}
}

@ARTICLE{2025A&A...700A..16T,
       author = {{Traianou}, E. and {G{\'o}mez}, J.~L. and {Cho}, I. and {Chael}, A. and {Fuentes}, A. and {Myserlis}, I. and {Wielgus}, M. and {Zhao}, G.-Y. and {Lico}, R. and {Moriyama}, K. and {Dey}, L. and {Bruni}, G. and {Dahale}, R. and {Toscano}, T. and {Gurvits}, L.~I. and {Lisakov}, M.~M. and {Kovalev}, Y.~Y. and {Lobanov}, A.~P. and {Pushkarev}, A.~B. and {Sokolovsky}, K.~V.},
        title = "{Revealing a ribbon-like jet in OJ 287 with RadioAstron}",
      journal = {\aap},
         year = 2025,
        month = aug,
       volume = {700},
          eid = {A16},
        pages = {A16},
          doi = {10.1051/0004-6361/202554929},
archivePrefix = {arXiv},
       eprint = {2508.01747},
 primaryClass = {astro-ph.HE}
}

@ARTICLE{2005astro.ph..3225L,
       author = {{Lobanov}, A.~P.},
        title = "{Resolution limits in astronomical images}",
      journal = {arXiv e-prints, arXiv:astro-ph/0503225},
         year = 2005,
        month = mar,
          doi = {10.48550/arXiv.astro-ph/0503225},
archivePrefix = {arXiv},
       eprint = {astro-ph/0503225},
 primaryClass = {astro-ph}
}

@article{vanbreugel1984,
  author  = {{van Breugel}, W. and {Miley}, G. and {Heckman}, T.},
  title   = {Studies of kiloparsec-scale steep-spectrum radio sources.
             {I.} {Multi}-frequency observations},
  journal = {\aj},
  year    = {1984},
  volume  = {89},
  pages   = {5--22},
  doi     = {10.1086/113480}
}

@article{planck2018,
  author  = {{Planck Collaboration} and {Aghanim}, N. and others},
  title   = {{Planck} 2018 results. {VI}. Cosmological parameters},
  journal = {\aap},
  year    = {2020},
  volume  = {641},
  pages   = {A6},
  doi     = {10.1051/0004-6361/201833910}
}

@article{Rodriguez_2006,
  author  = {Rodriguez, C. and Taylor, G. B. and Zavala, R. T. and Peck, A. B. and Pollack, L. K. and Romani, R. W.},
  title   = {A Compact Supermassive Binary Black Hole System},
  journal = {\apj},
  year    = {2006},
  volume  = {646},
  number  = {1},
  pages   = {49},
  doi     = {10.1086/504825}
}

@ARTICLE{2025A&A...695A.169P,
       author = {{P{\"o}tzl}, F.~M. and {Casadio}, C. and {Kalaitzidakis}, G. and {{\'A}lvarez-Ortega}, D. and {Kumar}, A. and {Missaglia}, V. and {Blinov}, D. and {Janssen}, M. and {Loudas}, N. and {Pavlidou}, V. and {Readhead}, A.~C.~S. and {Tassis}, K. and {Wilkinson}, P.~N. and {Zensus}, J.~A.},
        title = "{SMILE: Discriminating milli-lens systems in a VLBI pilot project}",
      journal = {\aap},
         year = 2025,
        month = mar,
       volume = {695},
          eid = {A169},
        pages = {A169},
          doi = {10.1051/0004-6361/202452340},
archivePrefix = {arXiv},
       eprint = {2409.15229},
 primaryClass = {astro-ph.GA}
}

@ARTICLE{2021MNRAS.507L...6C,
       author = {{Casadio}, C. and {Blinov}, D. and {Readhead}, A.~C.~S. and {Browne}, I.~W.~A. and {Wilkinson}, P.~N. and {Hovatta}, T. and {Mandarakas}, N. and {Pavlidou}, V. and {Tassis}, K. and {Vedantham}, H.~K. and {Zensus}, J.~A. and {Diamantopoulos}, V. and {Dolapsaki}, K.~E. and {Gkimisi}, K. and {Kalaitzidakis}, G. and {Mastorakis}, M. and {Nikolaou}, K. and {Ntormousi}, E. and {Pelgrims}, V. and {Psarras}, K.},
        title = "{SMILE: Search for MIlli-LEnses}",
      journal = {\mnras},
         year = 2021,
        month = oct,
       volume = {507},
       number = {1},
        pages = {L6-L10},
          doi = {10.1093/mnrasl/slab082},
archivePrefix = {arXiv},
       eprint = {2107.06896},
 primaryClass = {astro-ph.CO}
}

@ARTICLE{2009AJ....138.1874L,
       author = {{Lister}, M.~L. and {Cohen}, M.~H. and {Homan}, D.~C. and {Kadler}, M. and {Kellermann}, K.~I. and {Kovalev}, Y.~Y. and {Ros}, E. and {Savolainen}, T. and {Zensus}, J.~A.},
        title = "{MOJAVE: Monitoring of Jets in Active Galactic Nuclei with VLBA Experiments. VI. Kinematics Analysis of a Complete Sample of Blazar Jets}",
      journal = {\aj},
         year = 2009,
        month = dec,
       volume = {138},
       number = {6},
        pages = {1874-1892},
          doi = {10.1088/0004-6256/138/6/1874},
archivePrefix = {arXiv},
       eprint = {0909.5100},
 primaryClass = {astro-ph.CO}
}

@INPROCEEDINGS{1999ASPC..180..301F,
       author = {{Fomalont}, Ed B.},
        title = "{Image Analysis}",
    booktitle = {Synthesis Imaging in Radio Astronomy II},
         year = 1999,
       editor = {{Taylor}, G.~B. and {Carilli}, C.~L. and {Perley}, R.~A.},
       series = {ASP Conf. Ser.},
       volume = {180},
        month = jan,
        pages = {301}
}

@ARTICLE{2013MNRAS.435.3353H,
       author = {{Harwood}, Jeremy J. and {Hardcastle}, Martin J. and {Croston}, Judith H. and {Goodger}, Joanna L.},
        title = "{Spectral ageing in the lobes of FR-II radio galaxies: new methods of analysis for broad-band radio data}",
      journal = {\mnras},
         year = 2013,
        month = nov,
       volume = {435},
       number = {4},
        pages = {3353-3375},
          doi = {10.1093/mnras/stt1526},
archivePrefix = {arXiv},
       eprint = {1308.4137},
 primaryClass = {astro-ph.CO}
}

@ARTICLE{2003PASA...20...19M,
       author = {{Murgia}, Matteo},
        title = "{Spectral Ages of CSOs and CSS Sources}",
      journal = {\pasa},
         year = 2003,
        month = jan,
       volume = {20},
       number = {1},
        pages = {19-24},
          doi = {10.1071/AS02033},
archivePrefix = {arXiv},
       eprint = {astro-ph/0302376},
 primaryClass = {astro-ph}
}

@ARTICLE{2012A&A...537A..70S,
       author = {{Schinzel}, F.~K. and {Lobanov}, A.~P. and {Taylor}, G.~B. and {Jorstad}, S.~G. and {Marscher}, A.~P. and {Zensus}, J.~A.},
        title = "{Relativistic outflow drives {\ensuremath{\gamma}}-ray emission in 3C 345}",
      journal = {\aap},
         year = 2012,
        month = jan,
       volume = {537},
          eid = {A70},
        pages = {A70},
          doi = {10.1051/0004-6361/201117705},
archivePrefix = {arXiv},
       eprint = {1111.2045},
 primaryClass = {astro-ph.CO}
}

@ARTICLE{2010MNRAS.401.1709V,
       author = {{Vardoulaki}, Eleni and {Rawlings}, Steve and {Hill}, Gary J. and {Mauch}, Tom and {Inskip}, Katherine J. and {Riley}, Julia and {Brand}, Kate and {Croft}, Steve and {Willott}, Chris J.},
        title = "{The TexOx-1000 redshift survey of radio sources I: the TOOT00 region}",
      journal = {\mnras},
         year = 2010,
        month = jan,
       volume = {401},
       number = {3},
        pages = {1709-1759},
          doi = {10.1111/j.1365-2966.2009.15810.x},
archivePrefix = {arXiv},
       eprint = {0909.5691},
 primaryClass = {astro-ph.CO}
}

@article{petrov2021,
  author  = {{Petrov}, L.},
  title   = {The {RF}\/C Catalogue of {\sim}17000 Radio Sources},
  journal = {\aj},
  year    = {2021},
  volume  = {161},
  pages   = {14},
  doi     = {10.3847/1538-3881/abc4e1}
}

@incollection{greisen2003,
  author    = {{Greisen}, E.~W.},
  title     = {{AIPS}, the {VLA}, and the {VLBA}},
  booktitle = {Information Handling in Astronomy -- Historical Vistas},
  editor    = {{Heck}, A.},
  series    = {ASSL},
  volume    = {285},
  pages     = {109--125},
  year      = {2003},
  publisher = {Springer},
  doi       = {10.1007/0-306-48080-8_7}
}

@article{shepherd1994,
  author    = {{Shepherd}, M.~C. and {Pearson}, T.~J. and {Taylor}, G.~B.},
  title     = {{DIFMAP}: An interactive program for synthesis imaging},
  journal   = {BAAS},
  year      = {1994},
  volume    = {26},
  pages     = {987--989}
}

@article{hogbom1974,
  author  = {{H{\"o}gbom}, J.~A.},
  title   = {Aperture Synthesis with a Non-Regular Distribution
             of Interferometer Baselines},
  journal = {\aaps},
  year    = {1974},
  volume  = {15},
  pages   = {417}
}

@inproceedings{pearson1995,
  author    = {{Pearson}, T.~J.},
  title     = {Non-Imaging Data Analysis},
  booktitle = {Very Long Baseline Interferometry and the {VLBA}},
  editor    = {{Zensus}, J.~A. and {Diamond}, P.~J. and {Napier}, P.~J.},
  series    = {ASP Conf. Ser.},
  volume    = {82},
  pages     = {267},
  year      = {1995},
  publisher = {Astronomical Society of the Pacific}
}

@article{dow2009,
  author  = {{Dow}, J.~M. and {Neilan}, R.~E. and {Rizos}, C.},
  title   = {The International {GNSS} Service in a changing
             landscape of Global Navigation Satellite Systems},
  journal = {J. Geod.},
  year    = {2009},
  volume  = {83},
  pages   = {191--198},
  doi     = {10.1007/s00190-008-0300-3}
}

@article{noll2010,
  author  = {{Noll}, C.~E.},
  title   = {The Crustal Dynamics Data Information System:
             A resource to support scientific analysis using
             space geodesy},
  journal = {Adv. Space Res.},
  year    = {2010},
  volume  = {45},
  number  = {12},
  pages   = {1421--1440},
  doi     = {10.1016/j.asr.2010.01.018}
}

@misc{igs_ionex,
  author       = {{International GNSS Service}},
  title        = {{GNSS} Final Daily Ionosphere Total Electron
                  Content Grid Product},
  year         = {2022},
  publisher    = {NASA Crustal Dynamics Data Information System ({CDDIS})},
  address      = {Greenbelt, MD, USA},
  doi          = {10.5067/GNSS/gnss_igsionotec_001}
}

@ARTICLE{Agazie2023,
       author = {{Agazie}, Gabriella and {Anumarlapudi}, Akash and {Archibald}, Anne M. and {Arzoumanian}, Zaven and {Baker}, Paul T. and {B{\'e}csy}, Bence and {Blecha}, Laura and {Brazier}, Adam and {Brook}, Paul R. and {Burke-Spolaor}, Sarah and {Burnette}, Rand and {Case}, Robin and {Charisi}, Maria and {Chatterjee}, Shami and {Chatziioannou}, Katerina and {Cheeseboro}, Belinda D. and {Chen}, Siyuan and {Cohen}, Tyler and {Cordes}, James M. and {Cornish}, Neil J. and {Crawford}, Fronefield and {Cromartie}, H. Thankful and {Crowter}, Kathryn and {Cutler}, Curt J. and {Decesar}, Megan E. and {Degan}, Dallas and {Demorest}, Paul B. and {Deng}, Heling and {Dolch}, Timothy and {Drachler}, Brendan and {Ellis}, Justin A. and {Ferrara}, Elizabeth C. and {Fiore}, William and {Fonseca}, Emmanuel and {Freedman}, Gabriel E. and {Garver-Daniels}, Nate and {Gentile}, Peter A. and {Gersbach}, Kyle A. and {Glaser}, Joseph and {Good}, Deborah C. and {G{\"u}ltekin}, Kayhan and {Hazboun}, Jeffrey S. and {Hourihane}, Sophie and {Islo}, Kristina and {Jennings}, Ross J. and {Johnson}, Aaron D. and {Jones}, Megan L. and {Kaiser}, Andrew R. and {Kaplan}, David L. and {Kelley}, Luke Zoltan and {Kerr}, Matthew and {Key}, Joey S. and {Klein}, Tonia C. and {Laal}, Nima and {Lam}, Michael T. and {Lamb}, William G. and {Lazio}, T. Joseph W. and {Lewandowska}, Natalia and {Littenberg}, Tyson B. and {Liu}, Tingting and {Lommen}, Andrea and {Lorimer}, Duncan R. and {Luo}, Jing and {Lynch}, Ryan S. and {Ma}, Chung-Pei and {Madison}, Dustin R. and {Mattson}, Margaret A. and {McEwen}, Alexander and {McKee}, James W. and {McLaughlin}, Maura A. and {McMann}, Natasha and {Meyers}, Bradley W. and {Meyers}, Patrick M. and {Mingarelli}, Chiara M.~F. and {Mitridate}, Andrea and {Natarajan}, Priyamvada and {Ng}, Cherry and {Nice}, David J. and {Ocker}, Stella Koch and {Olum}, Ken D. and {Pennucci}, Timothy T. and {Perera}, Benetge B.~P. and {Petrov}, Polina and {Pol}, Nihan S. and {Radovan}, Henri A. and {Ransom}, Scott M. and {Ray}, Paul S. and {Romano}, Joseph D. and {Sardesai}, Shashwat C. and {Schmiedekamp}, Ann and {Schmiedekamp}, Carl and {Schmitz}, Kai and {Schult}, Levi and {Shapiro-Albert}, Brent J. and {Siemens}, Xavier and {Simon}, Joseph and {Siwek}, Magdalena S. and {Stairs}, Ingrid H. and {Stinebring}, Daniel R. and {Stovall}, Kevin and {Sun}, Jerry P. and {Susobhanan}, Abhimanyu and {Swiggum}, Joseph K. and {Taylor}, Jacob and {Taylor}, Stephen R. and {Turner}, Jacob E. and {Unal}, Caner and {Vallisneri}, Michele and {van Haasteren}, Rutger and {Vigeland}, Sarah J. and {Wahl}, Haley M. and {Wang}, Qiaohong and {Witt}, Caitlin A. and {Young}, Olivia and {Nanograv Collaboration}},
        title = "{The NANOGrav 15 yr Data Set: Evidence for a Gravitational-wave Background}",
      journal = {\apjl},
         year = 2023,
        month = jul,
       volume = {951},
       number = {1},
          eid = {L8},
        pages = {L8},
          doi = {10.3847/2041-8213/acdac6},
archivePrefix = {arXiv},
       eprint = {2306.16213},
 primaryClass = {astro-ph.HE}
}

@ARTICLE{1980ARA&A..18..165M,
       author = {{Miley}, G.},
        title = "{The structure of extended extragalactic radio sources}",
      journal = {\araa},
         year = 1980,
        month = jan,
       volume = {18},
        pages = {165-218},
          doi = {10.1146/annurev.aa.18.090180.001121}
}

@ARTICLE{2005ApJ...622..797K,
       author = {{Kataoka}, Jun and {Stawarz}, {\L}ukasz},
        title = "{X-Ray Emission Properties of Large-Scale Jets, Hot Spots, and Lobes in Active Galactic Nuclei}",
      journal = {\apj},
         year = 2005,
        month = apr,
       volume = {622},
       number = {2},
        pages = {797-810},
          doi = {10.1086/428083},
archivePrefix = {arXiv},
       eprint = {astro-ph/0411042},
 primaryClass = {astro-ph}
}

@article{Nyland_2026,
  author  = {Nyland, Kristina and Barrett, Mary Rachelle and Crom, Genna and Patil, Pallavi and Polisensky, Emil and Peters, Wendy and Giacintucci, Simona and Clarke, Tracy and Lacy, Mark and Mukundan, Shyaam and Dong, Dillon Z. and Goulding, Andy and Kimball, Amy E. and Kunert-Bajraszewska, Magdalena},
  title   = {A Compact Symmetric Object Discovered by the VLA Low-band Ionosphere and Transient Experiment},
  journal = {\apj},
  year    = {2026},
  volume  = {998},
  number  = {1},
  pages   = {168},
  doi     = {10.3847/1538-4357/ae0e12}
}

@ARTICLE{1983ApJ...264..296M,
       author = {{Marscher}, A.~P.},
        title = "{Accurate formula for the self-Compton X-ray flux density from a uniform, spherical, compact radio source.}",
      journal = {\apj},
         year = 1983,
        month = jan,
       volume = {264},
        pages = {296-297},
          doi = {10.1086/160597}
}

@ARTICLE{2008A&A...487..885O,
       author = {{Orienti}, M. and {Dallacasa}, D.},
        title = "{Are young radio sources in equipartition?}",
      journal = {\aap},
         year = 2008,
        month = sep,
       volume = {487},
       number = {3},
        pages = {885-894},
          doi = {10.1051/0004-6361:200809948},
archivePrefix = {arXiv},
       eprint = {0806.4831},
 primaryClass = {astro-ph}
}

@article{Panajyan1998,
  author  = {{Panajyan}, V.~G.},
  title   = {A new sample of gigahertz-peaked-spectrum, extra-galactic radio sources},
  journal = {Astrophysics},
  year    = {1998},
  volume  = {41},
  pages   = {246--253}
}

@ARTICLE{1997AJ....113..148O,
       author = {{O'Dea}, Christopher P. and {Baum}, Stefi A.},
        title = "{Constraints on Radio Source Evolution from the Compact Steep Spectrum and GHz Peaked Spectrum Radio Sources}",
      journal = {\aj},
         year = 1997,
        month = jan,
       volume = {113},
        pages = {148-161},
          doi = {10.1086/118241}
}

@article{Stanghellini2025,
  author  = {{Stanghellini}, C. and {Orienti}, M. and {Spingola}, C. and {Zanichelli}, A. and {Dallacasa}, D. and {Cassaro}, P. and {O'Dea}, C.~P. and {Baum}, S.~A. and {P{\'e}rez-Torres}, M.},
  title   = {Jetted subgalactic-size radio sources in merging galaxies. A jet redirection scenario},
  journal = {\aap},
  year    = {2025},
  volume  = {695},
  pages   = {A7},
  doi     = {10.1051/0004-6361/202451334},
  eprint  = {2407.02029},
  archivePrefix = {arXiv},
  primaryClass = {astro-ph.GA}
}

@article{KunertBajraszewska2010,
  author  = {{Kunert-Bajraszewska}, M. and {Gawro{\'n}ski}, M.~P. and {Labiano}, A. and {Siemiginowska}, A.},
  title   = {A survey of low-luminosity compact sources and its implication for the evolution of radio-loud active galactic nuclei. I. Radio data},
  journal = {\mnras},
  year    = {2010},
  volume  = {408},
  pages   = {2261--2278},
  doi     = {10.1111/j.1365-2966.2010.17271.x},
  eprint  = {1009.5235},
  archivePrefix = {arXiv},
  primaryClass = {astro-ph.CO}
}

@article{Orienti2012,
  author  = {{Orienti}, M. and {Dallacasa}, D.},
  title   = {A dying compact radio source in the local Universe: the case of B2 0258+35},
  journal = {\mnras},
  year    = {2012},
  volume  = {424},
  pages   = {532--544},
  doi     = {10.1111/j.1365-2966.2012.21218.x}
}

@article{Morganti2017,
  author  = {{Morganti}, R.},
  title   = {The many routes to AGN feedback},
  journal = {Nat. Astron.},
  year    = {2017},
  volume  = {1},
  pages   = {596--605},
  doi     = {10.1038/s41550-017-0223-0}
}

@article{Orienti2023,
  author  = {{Orienti}, M. and {Murgia}, M. and {Dallacasa}, D. and {Migliori}, G. and {D'Ammando}, F.},
  title   = {Young but fading radio sources: searching for remnants among compact steep-spectrum radio sources},
  journal = {\mnras},
  year    = {2023},
  volume  = {522},
  pages   = {3877--3891},
  doi     = {10.1093/mnras/stad1227},
  eprint  = {2304.12394},
  archivePrefix = {arXiv},
  primaryClass = {astro-ph.GA}
}

@article{Shimwell2022,
  author  = {{Shimwell}, T.~W. and {Hardcastle}, M.~J. and {Tasse}, C. and others},
  title   = {The LOFAR Two-metre Sky Survey. V. Second data release},
  journal = {\aap},
  year    = {2022},
  volume  = {659},
  pages   = {A1},
  doi     = {10.1051/0004-6361/202142484}
}

@article{Intema2017,
  author  = {{Intema}, H.~T. and {Jagannathan}, P. and {Mooley}, K.~P. and {Frail}, D.~A.},
  title   = {The GMRT 150 MHz all-sky radio survey. First alternative data release TGSS ADR1},
  journal = {\aap},
  year    = {2017},
  volume  = {598},
  pages   = {A78},
  doi     = {10.1051/0004-6361/201628536}
}

@article{Rengelink1997,
  author  = {{Rengelink}, R.~B. and {Tang}, Y. and {de Bruyn}, A.~G. and {Miley}, G.~K. and {Bremer}, M.~N. and {Roettgering}, H.~J.~A. and {Bremer}, M.~A.~R.},
  title   = {The Westerbork Northern Sky Survey (WENSS). I. A 570 square degree mini-survey around the north ecliptic pole},
  journal = {\aaps},
  year    = {1997},
  volume  = {124},
  pages   = {259--280},
  doi     = {10.1051/aas:1997358}
}

@article{Douglas1996,
  author  = {{Douglas}, J.~N. and {Bash}, F.~N. and {Bozyan}, F.~A. and {Torrence}, G.~W. and {Wolfe}, C.},
  title   = {The Texas Survey of Radio Sources Covering -35.5 degrees < declination < 71.5 degrees at 365 MHz},
  journal = {\aj},
  year    = {1996},
  volume  = {111},
  pages   = {1945},
  doi     = {10.1086/117932}
}

@article{Colla1973,
  author  = {{Colla}, G. and {Fanti}, C. and {Fanti}, R. and {Ficarra}, A. and {Formiggini}, L. and {Gandolfi}, E. and {Gioia}, I. and {Lari}, C. and {Marano}, B. and {Padrielli}, L. and {Tomasi}, P.},
  title   = {The B2 catalogue of radio sources. Third part},
  journal = {\aaps},
  year    = {1973},
  volume  = {11},
  pages   = {291}
}

@article{PaulinyToth1966,
  author  = {{Pauliny-Toth}, I.~I.~K. and {Wade}, C.~M. and {Heeschen}, D.~S.},
  title   = {Positions and Flux Densities of Radio Sources},
  journal = {\apjs},
  year    = {1966},
  volume  = {13},
  pages   = {65},
  doi     = {10.1086/190136}
}

@article{Condon1998,
  author  = {{Condon}, J.~J. and {Cotton}, W.~D. and {Greisen}, E.~W. and {Yin}, Q.~F. and {Perley}, R.~A. and {Taylor}, G.~B. and {Broderick}, J.~J.},
  title   = {The NRAO VLA Sky Survey},
  journal = {\aj},
  year    = {1998},
  volume  = {115},
  pages   = {1693--1716},
  doi     = {10.1086/300337}
}

@article{Gordon2021,
  author  = {{Gordon}, Y.~A. and {Boyce}, M.~M. and {O'Dea}, C.~P. and {Rudnick}, L. and {Andernach}, H. and {Vantyghem}, A.~N. and {Baum}, S.~A. and others},
  title   = {A Quick Look at the 3 GHz Radio Sky. I. Source Statistics from the Very Large Array Sky Survey},
  journal = {\apjs},
  year    = {2021},
  volume  = {255},
  pages   = {30},
  doi     = {10.3847/1538-4365/ac05c0},
  eprint  = {2102.11753},
  archivePrefix = {arXiv},
  primaryClass = {astro-ph.GA}
}

@article{Becker1991,
  author  = {{Becker}, R.~H. and {White}, R.~L. and {Edwards}, A.~L.},
  title   = {A New Catalog of 53522 4.85 GHz Sources},
  journal = {\apjs},
  year    = {1991},
  volume  = {75},
  pages   = {1},
  doi     = {10.1086/191529}
}

@article{Jackson2007,
  author  = {{Jackson}, N. and {Battye}, R.~A. and {Browne}, I.~W.~A. and {Joshi}, S. and {Muxlow}, T.~W.~B. and {Wilkinson}, P.~N.},
  title   = {A survey of polarization in the JVAS/CLASS flat-spectrum radio source surveys. I. The data and catalogue production},
  journal = {\mnras},
  year    = {2007},
  volume  = {376},
  pages   = {371--377},
  doi     = {10.1111/j.1365-2966.2007.11433.x},
  eprint  = {astro-ph/0703273},
  archivePrefix = {arXiv}
}

@ARTICLE{Kiehlmann2024,
  author  = {{Kiehlmann}, S. and {Readhead}, A. C. S. and {O'Neill}, S. and others},
  title   = {{Compact Symmetric Objects. II. Confirmation of a Distinct Population of High-luminosity Jetted Active Galaxies}},
  journal = {\apj}, year = 2024, volume = {961}, number = {2}, pages = {241},
  doi     = {10.3847/1538-4357/ad0cc2}
}

@ARTICLE{Orienti2010,
  author  = {{Orienti}, M. and {Murgia}, M. and {Dallacasa}, D.},
  title   = {{The last breath of the young gigahertz-peaked spectrum radio source PKS 1518+047}},
  journal = {\mnras}, year = 2010, volume = {402}, number = {3}, pages = {1892}
}

@ARTICLE{KunertBajraszewska2005,
  author  = {{Kunert-Bajraszewska}, M. and {Marecki}, A. and {Thomasson}, P. and {Spencer}, R.~E.},
  title   = {{FIRST-based survey of Compact Steep Spectrum sources. III.}},
  journal = {\aap}, year = 2005, volume = {440}, number = {1}, pages = {93}
}

@ARTICLE{Gugliucci2005,
  author  = {{Gugliucci}, N.~E. and {Taylor}, G.~B. and {Peck}, A.~B. and {Giroletti}, M.},
  title   = {{Dating COINS: Kinematic Ages for Compact Symmetric Objects}},
  journal = {\apj}, year = 2005, volume = {622}, number = {1}, pages = {136}
}

@ARTICLE{AnEtal2012,
       author = {{An}, T. and {Wu}, F. and {Yang}, J. and others},
        title = "{VLBI Observations of 10 Compact Symmetric Object Candidates: Expansion Velocities of Hot Spots}",
      journal = {\apjs},
         year = 2012,
        volume = {198},
        number = {1},
          eid = {5},
        pages = {5},
          doi = {10.1088/0067-0049/198/1/5}
}

@ARTICLE{An2026,
       author = {{An}, Tao},
        title = "{Time-domain Radio-loudness of Active Galactic Nuclei: Intermittency, Memory, and Jet Escape}",
      journal = {\apjl},
         year = 2026,
        volume = {1004},
          eid = {L5},
        pages = {L5},
          doi = {10.3847/2041-8213/ae6cde}
}

\begin{appendix}

\section{Modelfit parameters}
\label{app:table}

We summarize the best-fit model parameters in Table~\ref{tab:modelfit_txs0008}.

\begin{table*}[h!]
\centering
\caption{Modelfit parameters of \object{J0011+3443}.}
\label{tab:modelfit_txs0008}
\begin{tabular}{@{}lcccccccccccc@{}}
\hline\hline
\noalign{\smallskip}
Comp. & $\nu$ & $S$ & $\Delta S$ & $r$ & $\Delta r$ & $\theta_\mathrm{PA}$ &
$\Delta\theta_\mathrm{PA}$ & FWHM & $\Delta$FWHM & $T_\mathrm{b}$ &
$\Delta T_\mathrm{b}$ \\
 & (GHz) & (mJy) & (mJy) & (mas) & (mas) & ($\degr$) & ($\degr$) &
(mas) & (mas) & ($10^{9}$\,K) & ($10^{9}$\,K) \\
(1) & (2) & (3) & (4) & (5) & (6) & (7) & (8) & (9) & (10) & (11) & (12) \\
\noalign{\smallskip}
\hline
\noalign{\smallskip}
\multirow{4}{*}{A}
 & 2.3 & 128.0 & 12.8 & $\cdots$ & $\cdots$ & $\cdots$ & $\cdots$ & 3.42 & 0.34 & 4.86 & 1.08 \\
 & 4.9 & 106.3 & 10.6 & $\cdots$ & $\cdots$ & $\cdots$ & $\cdots$ & 1.27 & 0.13 & 6.30 & 1.44 \\
 & 8.5 & 28.0 & 2.8 & $\cdots$ & $\cdots$ & $\cdots$ & $\cdots$ & 0.82 & 0.08 & 1.31 & 0.29 \\
 & 23.6 & 1.9 & 0.2 & $\cdots$ & $\cdots$ & $\cdots$ & $\cdots$ & 0.36 & 0.04 & 0.06 & 0.01 \\
\hline
\multirow{4}{*}{B}
 &  2.3 &  67.4 &  6.7 & 39.86 & 1.20 & $-73.7$ & 1.7 & 2.88 & 0.29 & $>$3.61 & 0.81 \\
 &  4.9 &  58.5 &  5.9 & 39.31 & 1.18 & $-73.5$ & 1.7 & 0.95 & 0.10 & $>$6.20 & 1.44 \\
 &  8.5 &  14.2 &  1.4 & 39.21 & 1.18 & $-73.7$ & 1.7 & 0.50 & 0.05 & $>$1.78 & 0.40 \\
 & 23.6 &   1.0 &  0.1 & 38.96 & 1.17 & $-73.9$ & 1.7 & 0.32 & 0.03 & $>$0.04 & 0.01 \\
\hline
C & 23.6 &   0.5 &  0.1 &  0.64 & 0.06 & $+134.6$ & 5.3 & 0.26 & 0.03 & $>$0.03 & 0.01 \\
\hline
\end{tabular}
\tablefoot{Columns: (1) component ID, (2) observing frequency, (3) flux density, (4) flux-density uncertainty (10\% calibration error added in quadrature), (5) angular separation from A, (6) separation uncertainty, (7) position angle, (8) PA uncertainty, (9) Gaussian FWHM, (10) FWHM uncertainty, (11) brightness temperature, and (12) $T_\mathrm{b}$ uncertainty. For unresolved or marginally resolved components, the fitted FWHM is interpreted conservatively as an upper limit and the corresponding $T_\mathrm{b}$ as a lower limit.}
\end{table*}

\FloatBarrier

\section{uv-coverage}
\label{app:uvcov}

\begin{figure}[h!]
\centering
\includegraphics[width=\linewidth]{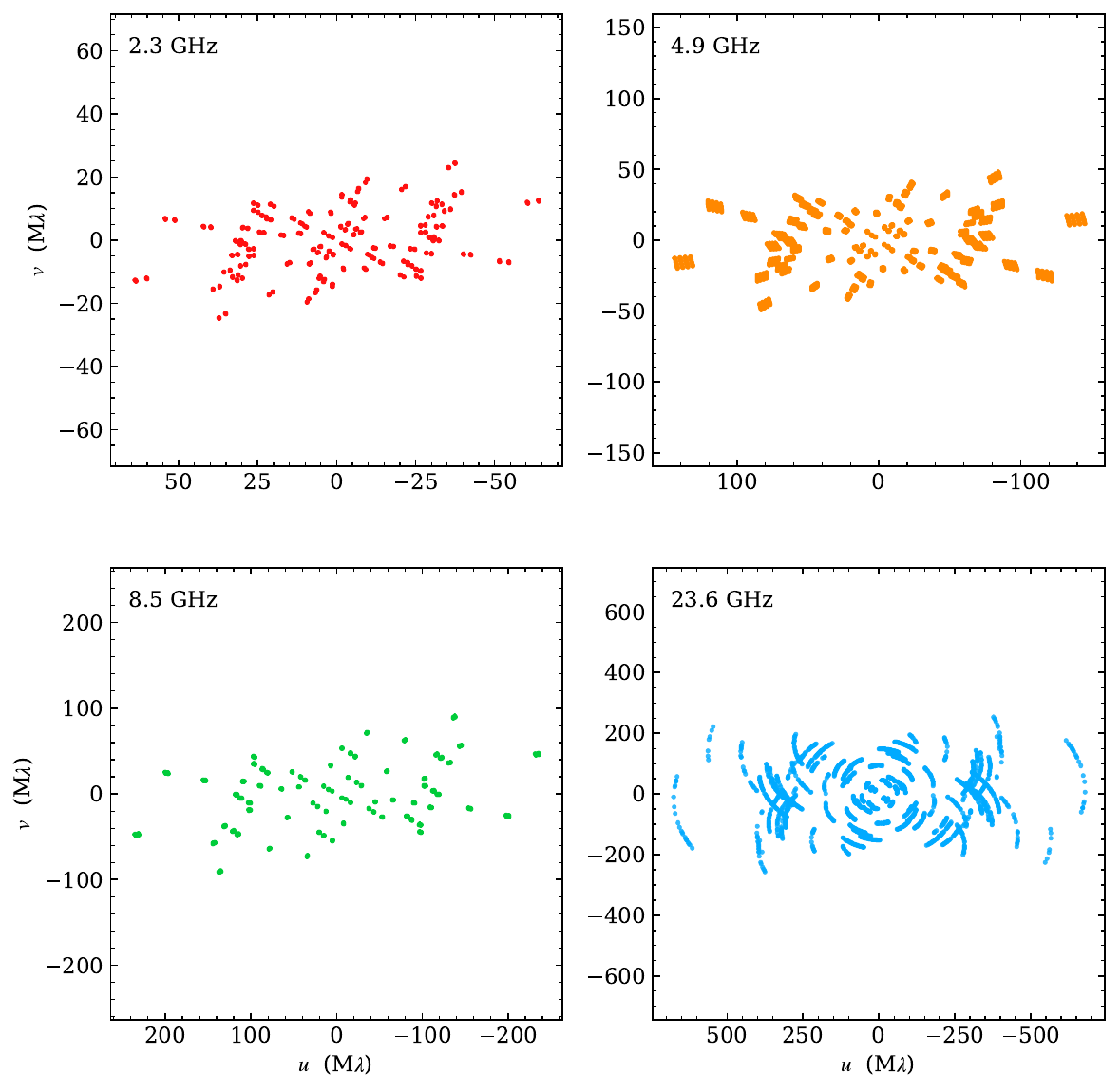}
\caption{Edited uv-coverage of the VLBA observations of
\object{J0011+3443}, obtained on 2022 March 8 at 2.3, 4.9, and
8.5\,GHz and on 2022 February 8 at 23.6\,GHz. Conjugate points are included. The projected baselines span approximately
$1.7$-$65$\,M$\lambda$ at 2.3\,GHz, $3.6$-$140$\,M$\lambda$ at 4.9\,GHz, $6.2$-$240$\,M$\lambda$ at 8.5\,GHz, and $17$-$640$\,M$\lambda$ at 23.6\,GHz. The shortest 2.3\,GHz spacing corresponds to a largest detectable angular scale of order $\sim120$\,mas, larger than the A-B separation, although diffuse low-surface-brightness emission may still be missed because of both the surface-brightness sensitivity limits and the sparse inner uv-coverage.}
\label{fig:uvcov}
\end{figure}

The uv-coverage of the edited data is shown in Fig.~\ref{fig:uvcov}. The shortest baselines are sensitive to angular scales larger than the A-B separation, but the absence of detected connecting emission does not exclude diffuse low-surface-brightness structure below the VLBA surface-brightness sensitivity.

\FloatBarrier

\section{Spectral curvature and equipartition magnetic field}
\label{app:aging}

The integrated low-resolution spectrum of \object{J0011+3443} steepens strongly above a few GHz (Sect.~\ref{sec:totalspec}). This steep GHz spectrum is consistent with an evolved electron population, but the present data do not provide a clean measurement of a synchrotron break frequency. The VLBA component spectra are affected by missing low-surface-brightness emission at some frequencies and are therefore not used to derive a source-wide cooling break.

\citet{2025A&A...695A.169P} measured EVN spectral indices $\alpha\approx-1.75$ between 4.3 and 7.6\,GHz for both components. The compact VLBA spectra in this work are not directly comparable to those EVN indices because the measurements have different frequency coverage, angular resolution, uv-coverage, and surface-brightness sensitivity. Differences between the measurements may reflect both spectral curvature and missing
flux, and do not establish a well-defined cooling break. We therefore do not derive a radiative age \citep{2003PASA...20...19M}. A robust spectral-age analysis would require multi-frequency observations that constrain the
optically thin spectral curvature, with matched angular resolution and comparable sensitivity to extended emission \citep[e.g.,][]{2013MNRAS.435.3353H}.

\subsection{Equipartition magnetic field}
\label{app:Beq}

Here we provide the details of the equipartition (minimum-energy) magnetic-field estimate summarized in Sect.~\ref{sec:discussion}. We use the formulation of \citet{2005ApJ...622..797K} \citep[see also][]{pacholczyk1970,1980ARA&A..18..165M} for a non-beamed synchrotron source, adopting a filling factor $\phi=1$, equal energy in relativistic protons and electrons ($k=1$), and no Doppler boosting ($\delta=1$), as appropriate for CSO-like lobe or hotspot emission. For each component we
use the 4.9-GHz compact VLBA flux density and angular size from Table~\ref{tab:modelfit_txs0008}, where the VLBA recovers a large fraction of the nearly frequency-matched low-resolution flux density
(Table~\ref{tab:recovered_flux}), obtaining

\begin{equation}
B_\mathrm{eq}^\mathrm{A}\approx11~\mathrm{mG}, \qquad
B_\mathrm{eq}^\mathrm{B}\approx12~\mathrm{mG},
\end{equation}

with an uncertainty of at least a factor of $\sim2$. These estimates refer to the compact VLBA-detected emission only; if additional diffuse low-surface-brightness lobe emission is present, the true source-averaged minimum-energy field may differ. The estimates are consistent with the
$\sim10$-$100$\,mG fields found in compact young radio sources and CSO/HFP hotspots \citep[e.g.,][]{2008A&A...487..885O}. They are also comparable to the equipartition field of $\sim10$\,mG reported for the CSO candidate \object{J0330-2730} at $z\approx0.90$ by \citet{Nyland_2026}.

A magnetic-field estimate based on homogeneous SSA requires reliable component turnover parameters and angular sizes \citep{1983ApJ...264..296M}. We do not derive such an estimate here because the individual VLBA component spectra are affected by missing flux. The integrated spectrum is comparably well described by the SSA-like and internal FFA parameterizations in Sect.~\ref{sec:totalspec}, so the present data do not uniquely determine the absorption
mechanism. Free-free absorption therefore remains a possible contributor. A similar ambiguity in the spectral modeling was reported for \object{J0330-2730} by \citet{Nyland_2026}, who found that both SSA and FFA models describe its peaked radio spectrum.

\end{appendix}

\end{document}